\documentclass[preprint,aps]{revtex4}
\usepackage{graphicx}
\usepackage{dcolumn}
\usepackage{bm}
\usepackage{color}
\usepackage{amsmath,amssymb,amsfonts}

\begin{document}

\title{Electromagnetic waves in uniaxial crystals: General formalism
with an application to Bessel beams}

\author{S. Hacyan}
\email{hacyan@fisica.unam.mx}
\author{R. J\'auregui}
\email{rocio@fisica.unam.mx}

\affiliation{  Instituto de F\'{\i}sica, Universidad Nacional
Aut\'onoma de M\'exico, Apdo. Postal 20-364, M\'exico D. F. 01000,
Mexico.}

\begin{abstract}
We present a mathematical formalism describing the propagation of a
completely general electromagnetic wave in a birefringent medium.
Analytic formulas for the refraction and reflection from a plane
interface are obtained. As a particular example, a Bessel beam
impinging at an arbitrary angle is analyzed in detail. Some
numerical results showing the formation and destruction of optical
vortices are presented.
\end{abstract}


 \maketitle

\section{Introduction}

The phenomena of birefringence was already well known at the time
when Huygens studied the ``strange properties of the Island crystal''
\cite{huygens} with a view to proving the wave nature of light.
Newton also discussed these same properties at length in his {\it
Opticks} \cite{newton}, but to prove precisely the contrary. In any
case, a full mathematical description of the phenomena turned out to
be quite cumbersome and it is only in the last few decades that
analytic expressions were obtained for the simplest case of a plane
electromagnetic wave in an uniaxial crystal \cite{lekner}.
Generalizations to more realistic waves, such as Hermite-Gauss,
Laguerre-Gauss  or Bessel beams, have been so far restricted to
paraxial (or almost paraxial) approximations \cite{fleck, martinez,
cia-cin-palma}, and to propagations parallel \cite{cia-cin-palma2} or
perpendicular \cite{cia-palma} to the crystal axis.

The aim of the present work is to obtain the most general
expressions describing reflected and refracted (ordinary and
extraordinary) waves in terms of the properties of a  beam impinging
at the plane interface of a uniaxial crystal. Analytic formulas are
presented in as compact a form as possible. As an example of
application, the propagation of a Bessel beam \cite{durnin, eberly}
is studied; these are electromagnetic modes that propagate in vacuum
with an invariant intensity pattern and exhibit polarization and
phase optical vortices \cite{nye} yielding an electromagnetic
orbital angular momentum (see, e. g., Ref. \cite{rozas}). However,
as we will show in the following, the intensity pattern inside a
birefringent crystal does not remain constant along the main axis of
propagation (unless it coincides with the crystal symmetry axis);
moreover optical vortices can be created or destroyed. Flossmann
{\it et al.} \cite{flossmann,flossmann2} considered the case of a
Laguerre-Gauss beam propagating inside a birefringent crystal and
analyzed the evolution of the polarization vortices in terms of
Stokes parameters. Here, we apply such a study to Bessel beams
using our analytical expressions, and complement it with a study of
the corresponding topological features of phase diagrams.

Modern experiments of parametric down-conversion use crystals that
are birefringent besides being nonlinear. These properties are
particularly important for generating entangled photons with
different dynamical properties determined by their source beam.
Thus, for example, when a Laguerre-Gauss or a Bessel beam is taken
as a pump, the down-converted light is expected to consist of
entangled photons with orbital angular momentum \cite{PDC}. However,
the anisotropy of the birefringent crystals used for that purpose
prevent a direct identification  of the expected characteristics of
the ordinary and extraordinary beams and, consequently, of the
properties of the idler and signal photons. Although these are
effects of quantum  nonlinear optics, their precise characterization
must be preceded by a detailed description of the classical linear
effects as we present in the following.

The organization of this article is as follows. In Section 2, a
formalism describing electromagnetic waves of any form inside and
outside a birefringent crystal is presented; the boundary conditions
at a plane interface are applied in order to obtain explicit
expressions for the reflected and refracted waves. The resulting
equations are used in the Section 3 to describe a Bessel beam
incident at an arbitrary angle on the crystal interface (though the
formalism is completely general, the optical axis is taken
perpendicular to the interface for simplicity); we show that the
reflected and refracted beams are given in terms of a single circuit
integral. Lastly, we discuss numerical results for some specific
parameters of the incident beam, following in particular the
evolution of optical vortices, both for polarization and for phase.

\section{Propagation in a birefringent medium.}

Consider an anisotropic medium described by a dielectric tensor
$\epsilon_{ij}$ or, alternatively, a dyad $\widehat{\epsilon}$ such
that the electric displacement is $ {\bf D}= \widehat \epsilon\cdot
{\bf E} $ and ${\bf B} = \mu {\bf H}$. For a birefringent medium,
$$
\widehat{\epsilon} = \epsilon {\bf 1} + \Delta \epsilon~{\bf s}~{\bf
s},
$$
where ${\bf s}$ is the axis of symmetry of the medium, $\Delta
\epsilon = \epsilon_{\parallel} - \epsilon$, and $\epsilon$ and
$\epsilon_{\parallel}$ are the permeabilities perpendicular and
parallel to the symmetry axis respectively.

The Maxwell equations in the absence of free charges and currents are
\begin{equation}
\nabla \cdot {\bf B} = 0~~, \quad \quad \nabla \times {\bf E} +
\frac{\partial {\bf B}}{\partial t}  =0~,
\end{equation}
\begin{equation}
\nabla \cdot {\bf D} = 0~~, \quad \quad \nabla \times {\bf H} -
\frac{\partial {\bf D }}{\partial t}  =0~.
\end{equation}
It is straightforward to see by direct substitution that their
general solution for a birefringent medium is
\begin{equation}
{\bf E}^O = {\bf s} \times \nabla \dot{\Psi}^O  ~, \quad ~~~~{\bf
E}^E = - \frac{1}{\epsilon} \nabla ({\bf s} \cdot \nabla \Psi^E) +
\mu \ddot{\Psi}^E ~{\bf s} \label{Etc}
\end{equation}
and
\begin{equation}
{\bf H}^O =  \frac{1}{\mu} \nabla ({\bf s} \cdot \nabla \Psi^O) -
\epsilon \ddot{\Psi}^O ~{\bf s}~, \quad ~~~~{\bf H}^E = {\bf s}
\times \nabla \dot{\Psi}^E~ \label{Htc}~,
\end{equation}
where ${\Psi}^O$ and ${\Psi}^E$ are Hertz potentials \cite{nisbet}
satisfying the two equations:
\begin{equation}
-\epsilon\mu \ddot{\Psi}^O + \nabla^2 \Psi^O =0 \label{psiO}
\end{equation}
and
\begin{equation}
-\epsilon_{\parallel}\epsilon\mu \ddot{\Psi}^E + \nabla \cdot
\widehat{\epsilon} \cdot \nabla \Psi^E =0 \label{psiE}.
\end{equation}
As it is well known, there are two fundamental modes: the ordinary
wave with ${\bf s} \cdot {\bf E}^O=0$ and the extraordinary wave with
${\bf s} \cdot {\bf B}^E=0$. Clearly, ${\Psi}^O$ and ${\Psi}^E$  are
associated to the ordinary and extraordinary waves respectively.

\subsection{Boundary conditions}

In the following, we restrict our analysis to harmonic fields of the
form $\Psi (t, {\bf x}) = e^{-i\omega t} \psi ({\bf x})$. Let the
vacuum be defined as the region $z < 0$ and consider a wave that
impinges from $z < 0$ and propagates inside the medium, in the region
$z > 0$, with wave vectors ${\bf k}^O = (k_x,k_y,k_z^O)$ and ${\bf
k}^E = (k_x,k_y,k_z^E)$, for the ordinary and extraordinary
components respectively.

The general solution of Eqs. (\ref{psiO}) and (\ref{psiE}) for $z>0$
is
\begin{equation}
\psi^{(O,E)}(x,y,z)
= \frac{1}{2\pi} \int ~dk_x dk_y~ e^{ik_x x +ik_y y+i k_z^{(O,E)}z}
~\widetilde{\psi}^{(O,E)}(k_x,k_y) ~\label{psiOE+},
\end{equation}
where $k_z^{(O,E)}$ in the integral must be taken as a function of
$k_x$ and $k_y$, namely $ k_z^O = (\epsilon\mu\omega^2 - k_x^2 -
k_y^2)^{1/2} \label{k_z}$, and $k_z^E$ as the solution for $k_z$ of
equation (\ref{ke}). Hereafter, the factor $e^{-i\omega t}$ is not
included for simplicity. The two-dimensional Fourier transforms in
the above formula are defined as
\begin{equation} \widetilde{\psi}^{(O,E)}(k_x,k_y)
=\frac{1}{2\pi} \int dx' ~dy'~ e^{-ik_x x' - ik_y y'}
~\psi^{(O,E)}(x',y',0^+).
\end{equation}

The (three-dimensional) Fourier transforms $\widetilde{{\bf E}}$ and
$\widetilde{{\bf H}}$ of the electric and magnetic fields follow from
Eqs. (\ref{Etc}) and (\ref{Htc}). Thus
\begin{equation}
\widetilde{{\bf E}}^O = \omega ~ {\bf s} \times {\bf k}^O~
\widetilde{\psi}^O ({\bf k}^O) ,\quad \widetilde{{\bf E}}^E =
\Big[\frac{1}{\epsilon} ({\bf s} \cdot {\bf k}^E){\bf k}^E~ - \mu
\omega^2  ~{\bf s}\Big]~\widetilde{\psi}^E ({\bf k}^E) \label{Eoe}~,
\end{equation}
and
\begin{equation}
\widetilde{{\bf H}}^O = \Big[-\frac{1}{\mu} ({\bf s} \cdot {\bf
k}^O){\bf k}^O~   + \epsilon \omega^2 ~{\bf
s}\Big]~\widetilde{\psi}^O ({\bf k}^O) ,\quad \widetilde{{\bf H}}^E =
\omega ~{\bf s} \times {\bf k}^E ~\widetilde{\psi}^E ({\bf k}^E)~,
\label{Hoe}
\end{equation}
in obvious notation, with the conditions
\begin{equation}
\epsilon \mu \omega^2 - ({\bf k}^O)^2 =0, \label{ko}
\end{equation}
\begin{equation}
\epsilon ~\epsilon_{\parallel} \mu ~\omega^2 - {\bf k}^E \cdot
\widehat{\epsilon} \cdot {\bf k}^E =0~.\label{ke}
\end{equation}

\subsection{Reflection and refraction}

In order to study the reflection and refraction of the wave, we
write the electric vector ${\bf E}$ in vacuum (that is, for $z<0$)
in the form
$$
{\bf E}(x,y,z) = \frac{1}{2\pi} \int  dk_x dk_y~ e^{ ik_x x +ik_y y}
$$
\begin{equation}
\Big[e^{i k_z z} ~\widetilde{{\bf E}}^I (k_x,k_y) + e^{-i k_z z}
~\widetilde{{\bf E}}^R (k_x,k_y) \Big]~,\label{EVIR}
\end{equation}
where now
\begin{equation}
k_z= (\omega^2 -k_x^2 -k_y^2)^{1/2}
\end{equation}
and $\widetilde{{\bf E}}^{(I,R)} (k_x,k_y)$ are the two-dimensional
Fourier transforms of the electric field components of the incident
and reflected waves, ${\bf E}^{(I,R)}(x,y,0^-)$ at the interface;
similar equations apply to the magnetic field component.

The boundary conditions imply the continuity of $E_x$, $E_y$, $H_x$
and $H_y$ at the interface $z=0$ (the continuity conditions on $D_z$
and $B_z$ are not independent since, from the Maxwell equations, $i
\omega D_z = \partial_y H_x - \partial_x H_y$ and $i \omega B_z =
\partial_x E_y - \partial_y E_x$). It is convenient to express each
Fourier transformed component of ${\bf B}$ and $E_z$ in the vacuum
region in terms of only $E_x$ and $E_y$ using the Maxwell equations.
For the incident field:
\begin{eqnarray}
&\widetilde{E}^I_z &= - ~\frac{1}{k_z} \Big(k_x \widetilde{E}^I_x +
k_y
\widetilde{E}^I_y \Big) \\
&\widetilde{B}^I_x &= - ~\frac{1}{k_z\omega} \Big[k_x k_y
\widetilde{E}^I_x + (k_y^2 + k_z^2)
\widetilde{E}^I_y \Big] \\
&\widetilde{B}^I_y &=  \frac{1}{k_z\omega} \Big[(k_x ^2 + k_z^2)
\widetilde{E}^I_x + k_x k_y \widetilde{E}^I_y \Big]\\
&\widetilde{B}^I_z &=  ~\frac{1}{\omega} \Big(-k_y \widetilde{E}^I_x
+ k_x \widetilde{E}^I_y \Big).
\end{eqnarray}
These equations can be rewritten in diad notation as
\begin{equation}
{\bf e}_z \times \widetilde{{\bf B}}^I = -k_z \omega ~(\omega^2
\widehat{1} - {\bf k}_{\bot} {\bf k}_{\bot})^{-1}~ \widetilde{{\bf
E}}^I_{\bot}
\end{equation}
(here and in the following, ${\bf V}_{\bot}=(V_x,V_y)$ for any vector
${\bf V}$).

For the reflected field, it is only necessary to change the sign of
$k_z$. Accordingly
\begin{equation}
{\bf e}_z \times (\widetilde{{\bf B}}^I + \widetilde{{\bf B}}^R) =
-k_z \omega ~(\omega^2 \widehat{1} - {\bf k}_{\bot} {\bf
k}_{\bot})^{-1}~ (\widetilde{{\bf E}}^I_{\bot} - \widetilde{{\bf
E}}^R_{\bot}) ~,
\end{equation}
and the boundary conditions take the form
\begin{equation}
\widetilde{{\bf E}}^I_{\bot} + \widetilde{{\bf E}}^R_{\bot}
=\widetilde{{\bf E}}^O_{\bot} + \widetilde{{\bf E}}^E_{\bot}~,
\end{equation}
and
\begin{equation}
\widetilde{{\bf E}}^I_{\bot} - \widetilde{{\bf E}}^R_{\bot} =
-~\frac{1}{k_z \omega}~ (\omega^2 \widehat{1} - {\bf k}_{\bot} {\bf
k}_{\bot}) ~[{\bf e}_z \times (\widetilde{{\bf H}}^O +
\widetilde{{\bf H}}^E)]~,
\end{equation}
where $\widetilde{{\bf E}}^O$, $\widetilde{{\bf E}}^E$,
$\widetilde{{\bf H}}^O$ and $\widetilde{{\bf H}}^O$ are to be taken
from (\ref{Eoe}) and (\ref{Hoe}). The above equations form a set of
four equations for the four unknown functions $\widetilde{E}^R_x$,
$\widetilde{E}^R_y$, $\widetilde{\psi}^O$ and $\widetilde{\psi}^E$ in
terms of $\widetilde{E}^I_x$ and $\widetilde{E}^I_y$. Explicitly we
have
\begin{eqnarray}
\widetilde{{\bf E}}^I_{\bot} + \widetilde{{\bf E}}^R_{\bot} &=& {\bf
P} ~\widetilde{\psi}^O +{\bf Q}~ \widetilde{\psi}^E \\ \nonumber
\widetilde{{\bf E}}^I_{\bot} - \widetilde{{\bf E}}^R_{\bot} &=& {\bf
R} ~\widetilde{\psi}^O +{\bf S} ~\widetilde{\psi}^E~,
\end{eqnarray}
where
\begin{equation}
{\bf P} =\omega ({\bf s} \times {\bf k}^O )_{\bot} ~,\quad {\bf Q} =
\epsilon^{-1}~ ({\bf s} \cdot {\bf k}^E)~ {\bf k}_{\bot} ~
 - \mu \omega^2   ~{\bf s}_{\bot}
\end{equation}
and
$$
{\bf R} =  \frac{\omega}{k_z} [\mu^{-1}~  ({\bf s} \cdot {\bf k}^O)~
{\bf e}_z \times {\bf k}_{\bot} ~  - \epsilon ~ \omega^2   ~{\bf
e}_z \times {\bf s} +\epsilon ~{\bf e}_z \cdot ({\bf s} \times {\bf
k}_{\bot})~{\bf k}_{\bot} ],
$$
\begin{equation}
{\bf S} = \frac{\omega^2}{k_z}(s_z {\bf k}_{\bot}- k_z^E {\bf
s}_{\bot}) + \frac{1}{k_z } [ ~({\bf s} \cdot {\bf k}_{\bot}) k_z^E -
s_z {\bf k}_{\bot}^2]~{\bf k}_{\bot}~.
\end{equation}

Therefore
\begin{equation}
\widetilde{\psi}^O = 2 ~\frac{[{\bf e}_z \times ({\bf Q} + {\bf S})]
\cdot \widetilde{{\bf E}}^I }{({\bf Q} + {\bf S}) \cdot [({\bf P} +
{\bf R}) \times {\bf e}_z ] }
\end{equation}
\begin{equation}
\widetilde{\psi}^E = -2 ~\frac{[{\bf e}_z \times ({\bf P} + {\bf
R})] \cdot \widetilde{{\bf E}}^I }{({\bf Q} + {\bf S}) \cdot [({\bf
P} + {\bf R}) \times {\bf e}_z ] }~,
\end{equation}
for the transmitted fields, and the full refracted electromagnetic
field is given for all its components by Eqs. (\ref{Eoe}) and
(\ref{Hoe}).

Also
\begin{equation}
\widetilde{{\bf E}}^R_{\bot} = \frac{1}{2} ({\bf P} - {\bf R})
\widetilde{\psi}^O +\frac{1}{2} ({\bf Q} - {\bf S})
\widetilde{\psi}^E~, \label{ER}
\end{equation}
\begin{equation}
\widetilde{{\bf E}}^R_z = ~\frac{1}{k_z} \Big(k_x \widetilde{E}^R_x +
k_y \widetilde{E}^R_y \Big) ,\label{ERz}
\end{equation}
for the reflected fields.

The electromagnetic potentials $\psi^{(O,E)}(x,y,z)$ can be obtained
by Fourier transforms from Eq.(\ref{psiOE+}), from where the
transmitted electric fields ${\bf E}^{(O,E)}$ are obtained with Eqs.
(\ref{Etc}) and (\ref{Htc}). Similarly, the reflected wave follows
from the Fourier transform of Eq. (\ref{ER}).

\subsection{Crystal axis perpendicular to interface}

The above equations can be solved for a general orientation of the
crystal axis, although the resulting expressions are somewhat
cumbersome. They do simplify considerably in the particular case of
the crystal axis perpendicular to the interface. Accordingly, if
${\bf s}={\bf e}_z$, then
\begin{equation}
\epsilon_{\parallel} \mu \omega^2 = k_x^2 + k_y^2 +
\frac{\epsilon_{\parallel}}{\epsilon} (k_z^E)^2
\end{equation}
and therefore
\begin{eqnarray}
{\bf Q} \pm {\bf S}&=& k_z \Big( \frac{k_z^E}{\epsilon k_z} \pm 1
\Big)~{\bf k}_{\bot} \\ \nonumber {\bf P} \pm {\bf R} &=&\omega
\Big( 1 \pm \frac{k_z^O}{\mu k_z} \Big)~{\bf e}_z \times {\bf
k}_{\bot},
\end{eqnarray}
from where it follows that
\begin{equation}
 \Big[ ({\bf Q} + {\bf S}) \times ~({\bf
P} + {\bf R})\Big]_z  = k_z k_{\bot}^2 \omega \Big(
\frac{k_z^E}{\epsilon k_z} + 1 \Big)~\Big(1 + \frac{k_z^O}{\mu k_z}
\Big) ~.
\end{equation}

Thus
\begin{equation} \psi^O(x,y,z)= \frac{1}{\pi} \int ~dk_x
dk_y~ e^{ik_x x +ik_y y+i k_z^O z} \frac{\mu
k_z ~ }{ k_{\bot}^2 ( \mu k_z + k_z^O) }\widetilde{B}^I_z
~,\label{psiperpO}
\end{equation}
\begin{equation}
\psi^E(x,y,z)= -\frac{1}{\pi} \int ~dk_x dk_y~ e^{ik_x x +ik_y y+i
k_z^E z}\frac{\epsilon k_z ~ }{k_{\bot}^2 (\epsilon k_z +
k_z^E) }\widetilde{E}^I_z ~,\label{psiperpE}
\end{equation}
and the electric field components of the ordinary and extraordinary
waves can be obtained from Eqs. (\ref{Etc}) and (\ref{Htc}). Notice
that we have used the Maxwell equations in order to substitute
$({\bf e}_z \times {\bf k}_{\bot}) \cdot \widetilde{{\bf E}}^I =
\omega \widetilde{B}^I_z$ and ${\bf k}_{\bot} \cdot  \widetilde{{\bf
E}}^I =-k_z\widetilde{E}^I_z$ in the above equations.

As for the reflected wave,
\begin{equation}
{\bf E}^R = \frac{1}{2\pi} \int \int dk_x dk_y~ e^{ik_x x +ik_y y-i
k_z z} ~\widetilde{{\bf E}}^R~,
\end{equation}
where $\widetilde{{\bf E}}^R$ is given by Eqs. (\ref{ER}) and
(\ref{ERz}).

\section{Bessel beams}

As an example, let us consider a Bessel beam that impinges at a given
angle onto the surface of the crystal. For simplicity, we consider
only the case of the crystal axis perpendicular to the interface. If
$\omega$ is the frequency and $\zeta$ is the axicon angle, then the incident
beam can be written in the general form:
$$
{\bf E}= \frac{1}{2 \pi \omega} \int d^3{\bf k}  ~e^{i{\bf k}\cdot
{\bf r}} \Big[ \frac{1}{2}({\cal E} {\rm cosec} \zeta + i {\cal B}
\cot \zeta) \Big( \frac{{\bf k}\cdot {\bf e}_p}{i|{\bf k} \cdot {\bf
e}_p|}\Big)^{m-1}  {\bf e}_p
$$
$$
+\frac{1}{2} ({\cal E} {\rm cosec} \zeta - i {\cal B} \cot \zeta)
\Big(\frac{{\bf k}\cdot {\bf e}_p}{i|{\bf k} \cdot {\bf
e}_p|}\Big)^{m+1}  {\bf e}_p^*
$$
\begin{equation}
+  {\cal B}  \Big(\frac{{\bf k}\cdot {\bf e}_p}{i|{\bf k} \cdot {\bf
e}_p|}\Big)^m {\bf e}_q ~ \Big] ~\delta ({\bf k} \cdot {\bf e}_q -
\omega \cos \zeta) ~\delta (|{\bf k}| - \omega ) ~, \label{bbeam}
\end{equation}
where ${\bf e}_q$ is the direction of propagation of the beam and
${\bf e}_p$ is the standard (non normalized) left hand polarization
vector perpendicular to ${\bf e}_q$. The magnetic field ${\bf B}$ is
given by the same expression as above with only the interchange
${\cal B} \rightarrow {\cal E}$ and ${\cal E} \rightarrow -{\cal
B}$.

Taking ${\bf e}_q= {\bf e}_z$ and  ${\bf e}_p = {\bf e}_x + i {\bf
e}_y$, we obtain:
\begin{eqnarray}
{\bf E}  = \frac{1}{2 k_\bot}
 e^{-i\omega t + i k_z z} &\Big[&(\omega {\cal E} + i k_z {\cal B})
J_{m-1} (k_\bot \rho) e^{i (m-1) \phi} (\hat{{\bf e}}_x + i
\hat{{\bf e}}_y) \nonumber\\ &+& (\omega {\cal E} -i k_z {\cal B} )
J_{m+1} (k_\bot \rho) e^{i (m+1) \phi} (\hat{{\bf e}}_x - i
\hat{{\bf e}}_y)~~\Big]\nonumber\\
&+& e^{-i\omega t + i k_z z} {\cal B} J_{m} (k_\bot \rho) e^{i m
\phi} \hat{{\bf e}}_z~, \label{1_a}
\end{eqnarray}
where $k_z =\omega\cos\zeta$ and $k_\bot = \omega\sin\zeta$, in
complete accordance with the standard expressions for Bessel beams
\cite{durnin} (see, {\it e. g.}, Eq. (2.6) of our previous paper
\cite{hj}).

In order to consider a beam impinging at an arbitrary angle $\alpha$,
we set
$${\bf e}_q= \sin \alpha ~{\bf e}_y + \cos \alpha ~{\bf e}_z$$ and
$${\bf e}_p = {\bf e}_x + i (\cos \alpha ~{\bf e}_y - \sin
\alpha ~{\bf e}_z),$$ without further loss of generality.

Due to the first Dirac delta function in Eq.~(\ref{bbeam}), the $k_z$
integration can be performed directly setting
\begin{equation}
k_z= (\omega \cos \zeta - k_y \sin
\alpha)/\cos \alpha
\end{equation}
in the integrand. Thus the two-dimensional Fourier transform of the
electric field at the interface takes the form
$$
\widetilde{E}_z (k_x,k_y)= \frac{1}{\omega } \Big[ -\frac{i}{2}\tan
\alpha ~({\cal E} {\rm cosec} \zeta + i {\cal B} \cot \zeta)
\Big(\frac{{\bf k}\cdot {\bf e}_p}{i|{\bf k} \cdot {\bf
e}_p|}\Big)^{m-1}
$$
\begin{equation}
+\frac{i}{2} \tan \alpha ~({\cal E} {\rm cosec} \zeta - i {\cal B}
\cot \zeta) \Big(\frac{{\bf k}\cdot {\bf e}_p}{i|{\bf k} \cdot {\bf
e}_p|}\Big)^{m+1} +  {\cal B}  \Big(\frac{{\bf k}\cdot {\bf
e}_p}{i|{\bf k} \cdot {\bf e}_p|}\Big)^m  ~ \Big] ~\delta (|{\bf k}
| - \omega ) ~,
\end{equation}
where now
$$
{\bf k}\cdot {\bf e}_p = k_x + i (k_y \sec \alpha - \omega \tan
\alpha \cos \zeta).
$$
The magnetic field component $\widetilde{B}_z (k_x,k_y)$ is obtained
from the above expression by simply changing ${\cal B} \rightarrow
{\cal E}$ and ${\cal E} \rightarrow -{\cal B}$.

The next step is to substitute the above expressions in
Eqs.~(\ref{psiperpO}) and (\ref{psiperpE}). In order to perform the
corresponding integral, the following change of variables to
ellipsoidal coordinates $(U,V)$ is appropriate:
\begin{eqnarray}
k_x &=& \omega~\sin \alpha ~\sin \zeta ~\cosh U \cos V \\ \nonumber
k_y &=& \omega ~ \sin \alpha ~[\cos \zeta  + \sin \zeta ~\sinh U
~\sin V].
\end{eqnarray}
It then follows after some straightforward algebra that
\begin{equation}
{\bf k}\cdot {\bf e}_p = \omega \sin \zeta \Big[ \Big( \frac{\cosh
U}{\cosh U_0} \Big) \cos V + i \Big( \frac{\sinh U}{\sinh U_0} \Big)
\sin V \Big],
\end{equation}
and
\begin{equation}
\delta (|{\bf k}| - \omega) ~dk_x ~dk_y = \omega ~\cos \alpha ~\delta
(U - U_0) ~dU ~dV~,
\end{equation}
where $ \tanh U_0 \equiv \cos \alpha$.

Summing up,
\begin{eqnarray}
k_x(V)&=&\omega \sin \zeta \cos V \\ \nonumber k_y(V)&=&\omega ~(\cos
\alpha~\sin \zeta ~ \sin V + \sin \alpha ~\cos \zeta ) \\ \nonumber
k_z(V) &=& \omega (- \sin \alpha \sin \zeta \sin V + \cos \alpha \cos
\zeta ),  \nonumber
\end{eqnarray}
$k^2_{\bot}(V) = k_x^2 (V) + k_y^2 (V)$ and
$$
k_z^O(V) = \sqrt{ \epsilon \mu \omega^2 -  k_{\bot}^2(V)},
$$
$$
k_z^E(V) = \sqrt{ \epsilon \mu \omega^2
-\frac{\epsilon}{\epsilon_{\parallel}}  k_{\bot}^2(V)}~.
$$

The final result for the ordinary and extraordinary waves can be
expressed as a circuit integral:
$$
{\bf E}^O ({\bf r}) = \frac{1}{\pi} i^{-m}  \int_0^{2\pi} dV~ e^{i
[{\bf r}\cdot {\bf k}^O(V) + m V]} \frac{\mu \omega k_z(V) ~
}{k_{\bot}^2(V) [ \mu k_z(V) + k_z^O(V)] }
$$
\begin{equation}
\times
 \Big[~ \frac{\sin \alpha}{\sin \zeta} ~(-{\cal B} ~  \cos V +
{\cal E} \cos \zeta \sin V) + {\cal E}\cos \alpha \Big]~({\bf e}_z
\times {\bf k}) ,\label{psiObessel}
\end{equation}
and
$$
{\bf E}^E ({\bf r}) =  \frac{1}{\pi} i^{-m} \int_0^{2\pi} dV~ e^{i
[{\bf r}\cdot {\bf k}^E(V) + m V]}\frac{ k_z(V) }{ k_{\bot}^2(V)
[\epsilon k_z(V) + k_z^E(V)] }
$$
\begin{equation}
\times \Big[ ~ \frac{\sin \alpha}{\sin \zeta} ~({\cal E} ~  \cos V +
{\cal B} \cos \zeta \sin V) + {\cal B} \cos \alpha \Big]~[ k_z^E (V)
{\bf k}^E (V) - \epsilon \mu \omega^2 {\bf e}_z].\label{psiEbessel}
\end{equation}

For the reflected wave:
$$
{\bf E}^R_{\bot} ({\bf r}) = \frac{1}{2\pi} i^{-m}  \int_0^{2\pi} dV~
e^{i [x k_x (V) + y k_y (V) - z k_z (V) + m V]}
\frac{1}{k_{\bot}^2(V)} \Big\{ \omega ~\frac{\mu k_z(V) - k_z^O(V)}{
\mu k_z(V) + k_z^O(V)}
$$
$$
\times \Big[~ \frac{\sin \alpha}{\sin \zeta} ~(-{\cal B} ~  \cos V +
{\cal E} \cos \zeta \sin V) + {\cal E}\cos \alpha \Big]({\bf e}_z
\times {\bf k})
$$
\begin{equation}
+ k_z(V) ~\frac{k_z^E(V) - \epsilon k_z(V)}{k_z^E(V) + \epsilon
k_z(V)} \Big[ ~ \frac{\sin \alpha}{\sin \zeta} ~({\cal E} ~  \cos V +
{\cal B} \cos \zeta \sin V) + {\cal B} \cos \alpha \Big]{\bf
k}_{\bot} \Big\}\label{psiRbessel}
\end{equation}
and
$$
E^R_z ({\bf r}) = \frac{1}{2\pi} i^{-m}  \int_0^{2\pi} dV~ e^{i [x k_x
(V) + y k_y (V) - z k_z (V) + m V]}
$$
\begin{equation}
~\frac{k_z^E(V) - \epsilon k_z(V)}{k_z^E(V) + \epsilon k_z(V)} \Big[ ~
\frac{\sin \alpha}{\sin \zeta} ~({\cal E} ~  \cos V + {\cal B} \cos
\zeta \sin V) + {\cal B} \cos \alpha \Big].\label{psiRzbessel}
\end{equation}
Notice that the reflected beam is invariant under propagation along the
main propagation axis. This axis makes an angle $-\alpha$ with the
normal of the interface surface.

For a beam impinging perpendicularly to the interface, $\alpha =0$,
the above integrals can be analytically solved in terms of Bessel
functions with a proper scaling of the parameter $\zeta$. The result
is
\begin{eqnarray}
{\bf E}^O  = &{\cal E}&\frac{\mu\omega \cos\zeta }{(\mu k_z +
k_z^O)\sin\zeta}
 e^{-i\omega t + i k_z^O z } \Big[
J_{m-1} (k_\bot \rho) e^{i (m-1) \phi} (\hat{{\bf e}}_x + i \hat{{\bf
e}}_y) \nonumber\\ &+&  J_{m+1} (k_\bot \rho) e^{i (m+1) \phi}
(\hat{{\bf e}}_x - i \hat{{\bf e}}_y)~~\Big] ~,\label{Eo_a}
\end{eqnarray}
\begin{eqnarray}
{\bf E}^E  = &{\cal B}&\frac{k_z^E~\cos\zeta }{(\epsilon k_z
+k_z^E)\sin\zeta}
 e^{-i\omega t + i k_z^E z} \Big[
J_{m-1} (k_\bot \rho) e^{i (m-1) \phi} (\hat{{\bf e}}_x + i
\hat{{\bf e}}_y) \nonumber\\ &+&  J_{m+1} (k_\bot \rho) e^{i (m+1)
\phi} (\hat{{\bf e}}_x - i
\hat{{\bf e}}_y)~~\Big]\nonumber\\
&-&\frac{2\omega\epsilon\sin\zeta}{\epsilon_{\|}} e^{-i\omega t + i
k_z^E z} J_{m} (k_\bot \rho) e^{i m \phi} \hat{{\bf e}}_z~,
\label{Ee_a}
\end{eqnarray}
where $k_z = \omega \cos \zeta$, $k_z^O = \omega \sqrt{\epsilon\mu -
\sin^2 \zeta}$ and $k_z^E = \omega \sqrt{\epsilon\mu -
(\epsilon/\epsilon_{\parallel})\sin^2 \zeta}$. These expressions show
explicitly the polarizing effect of birefringence.

\begin{figure}
\includegraphics[scale=0.3]{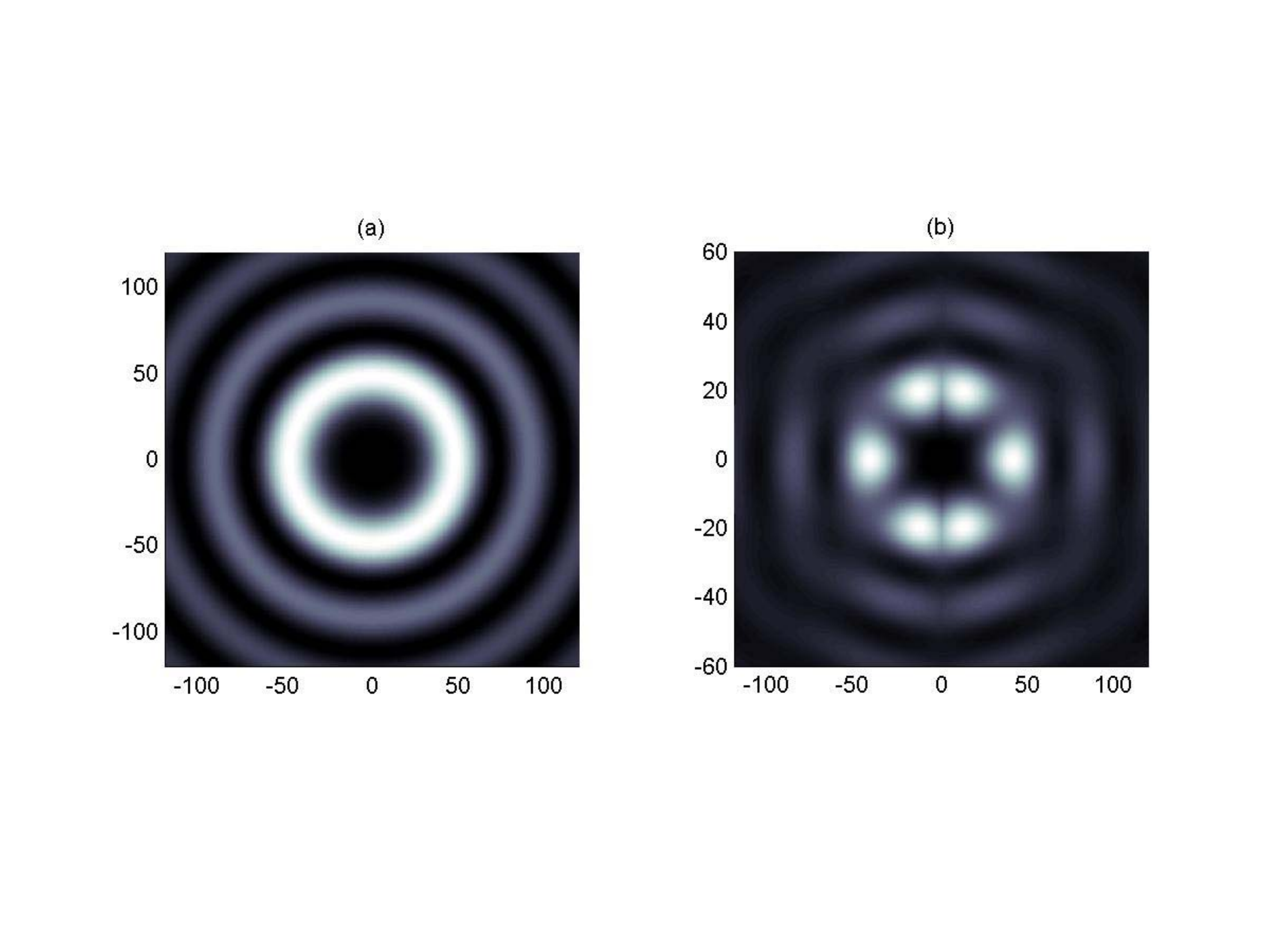}
\caption{Intensity pattern of an incident Bessel beam of order $m=2$
impinging a calcite crystal at an angle $\alpha =\pi/3$ as seen in a
plane (a) perpendicular to the main axis of propagation and (b)
parallel to the interface (notice the change of scale).  The  beam
is circularly polarized, ${\cal E} = i{\cal B}$, and $\zeta
=\pi/36$. The width of the observation window is measured in units
of $\lambda = c/\omega$}. \label{fig1}
\end{figure}

\begin{figure}
\includegraphics[scale=0.125]{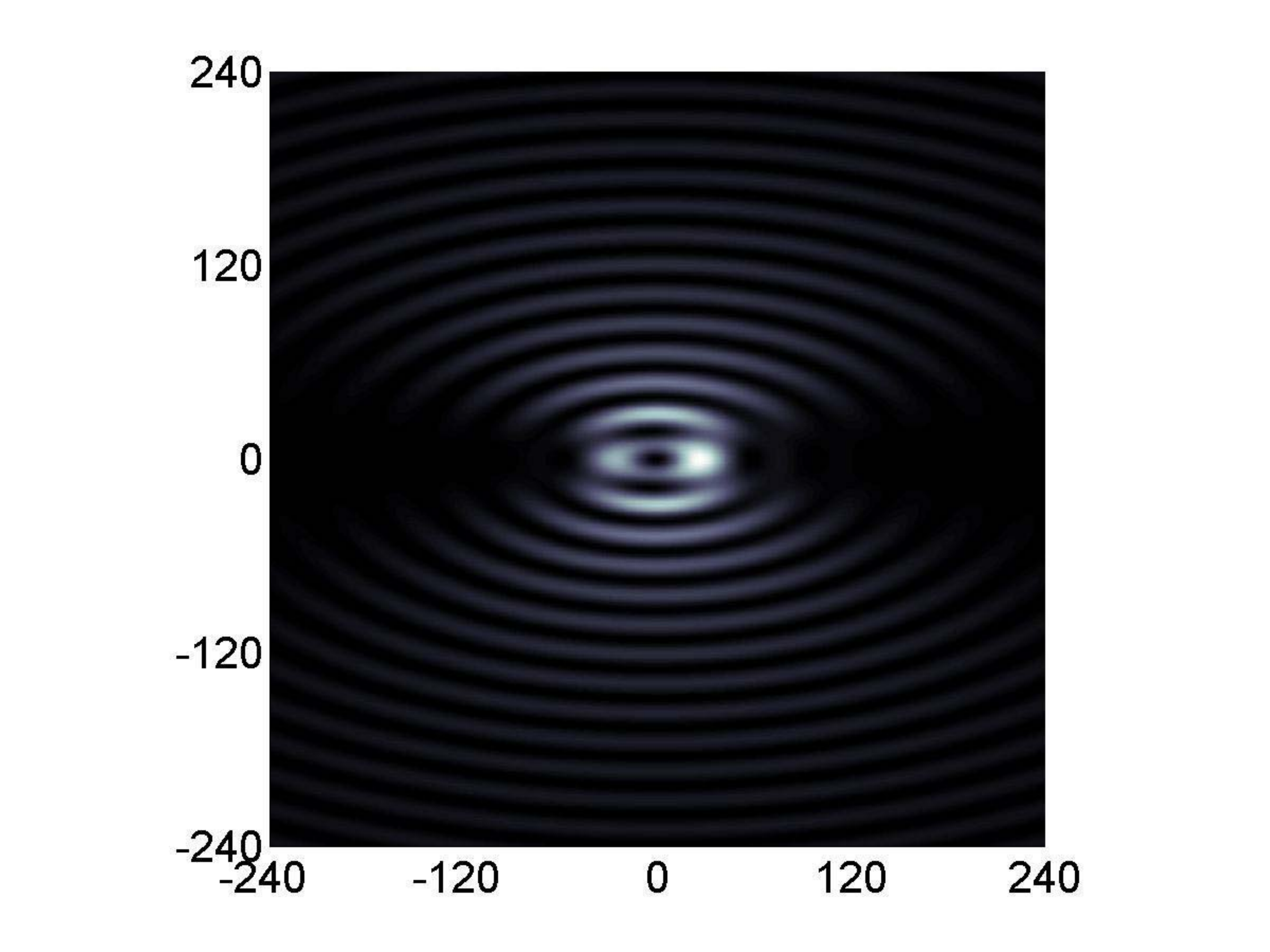}
\caption{Intensity pattern of the reflected beam resulting from an
incident Bessel beam of order $m=2$ with the same configuration as
described in Fig.~1. The plane of observation is tilted to be
perpendicular to the main direction of propagation. This beam has an
intensity pattern invariant under propagation along that axis. The
width of the observation window is measured in units of $\lambda =
c/\omega$.} \label{fig2}
\end{figure}

\begin{figure}
\includegraphics[scale=0.8]{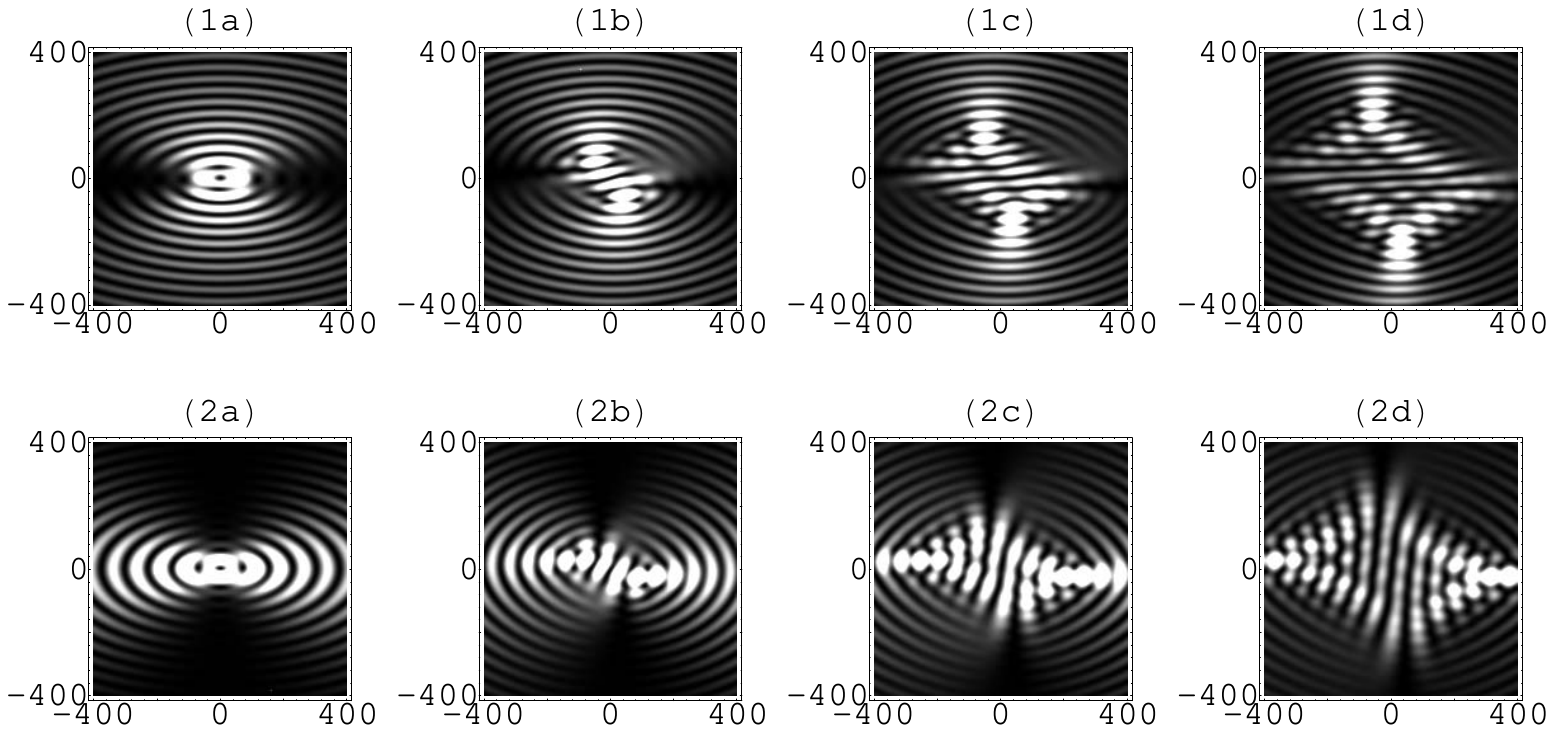}
\caption{Intensity pattern of the ordinary (1st row) and
extraordinary (2nd row) beams resulting from an incident Bessel beam
of order $m=2$ with the same configuration as described in Fig.~1.
The beam intensity pattern shows a richer topological structure as it
propagates. The width of the observation window is given in units of
$\lambda = c/\omega$. In Figures (a), the observation plane is the
interface itself; in the other figures, it is tilted to be
perpendicular to the main propagation axis and is located at:  (b)$ z
=3500$ $ \lambda$; (c)$ z =7000$ $ \lambda$; (d) $ z =10500$ $
\lambda$.} \label{fig3}
\end{figure}

Figure (\ref{fig1}) illustrates the intensity pattern of a Bessel
wave in a plane perpendicular to the mean direction of propagation
and in a plane parallel to the interface.  The corresponding
intensity patterns of the reflected, ordinary  and extraordinary
beams for a calcite crystal ($\epsilon$ = 2.748964,
$\epsilon_\parallel$ = 2.208196) are illustrated in
Figs.(\ref{fig2}-\ref{fig3}). In these examples, the second order
($m=2$) incident Bessel beam is taken at almost the paraxial limit
$\zeta =\pi/36$, the incidence angle is $\alpha = \pi/3$ and the
incident wave is a linear superposition of a transverse electric and
a transverse magnetic beam with equal amplitudes and a $\pi/2$ phase
difference: ${\cal E} = i{\cal B}$. The intensity pattern of the
reflected beam (Fig.~2), at the usual angle $\alpha_R=-\alpha$, is
similar to that of the incident one. As mentioned above, it is
invariant under propagation and exhibits an elliptic-like symmetry.
The ordinary and extraordinary waves are illustrated in Figs.~~(3);
for both refracted waves, it is possible to identify an axis of
symmetry, though the beams are not propagation invariant. A textbook
result is that for a plane wave, the refraction angles of the
ordinary and extraordinary waves, $\alpha^{PW}_O$ and $\alpha^{PW}_E$
respectively, are given in our case by
\begin{equation} \sin \alpha^{PW}_O = \frac{\sin
\alpha}{\sqrt{\epsilon\mu}}
\end{equation}
\begin{equation}
\sin \alpha^{PW}_E = \frac{\sin \alpha}{\sqrt{\epsilon\mu + (\Delta
\epsilon /\epsilon_{\parallel})\sin^2 \alpha}}~.
\end{equation}
However, for a Bessel beam, the incident wave is a superposition of
plane waves propagating in a cone with an aperture given by the
axicon angle $\zeta$; it is only in the limit $\zeta\rightarrow 0$
that the above equations define the correct angle for the axis of
symmetry $\alpha_O$ and $\alpha_E$. In general, no analytical
expression for an average value of $\alpha_O$ and $\alpha_E$ could be
found.

The intensity patterns (1a) and (2a) in Fig.~(3) are given at the
interface surface, and all the other patterns are evaluated on planes
perpendicular to the main propagation axis of each wave. The latter
were obtained performing the corresponding rotations for the
observation points and electric fields. The intensity patterns of
the ordinary and extraordinary beams have a structure similar to
that of the reflected wave near the interface, but this structure
gets gradually more complex as they propagate.

Since in this example the ordinary, extraordinary and reflected
electric fields are approximately contained in a plane perpendicular
to their main direction of propagation, the topological structure of
their polarization can be studied using Stokes parameters \cite{nye};
these are defined as
\begin{eqnarray}
S_0&=& \vert E_x\vert^2 + \vert E_y\vert^2\nonumber\\
S_1&=& \vert E_x\vert^2 - \vert E_y\vert^2\nonumber\\
S_2&=& E^*_xE_y + E_xE^*_y\nonumber\\
S_3&=&-i(E^*_xE_y - E_xE^*_y)~,\nonumber\\
\end{eqnarray}
and are given in terms of light intensities for different
orientations of an analyzer. The condition $S_3({\bf x})=0$
corresponds to local linear polarization at the point ${\bf x}$, and
$S_1({\bf x}) = S_2({\bf x}) =0$  to circular polarization. A
polarization singularity corresponds to $S_1({\bf x}) = S_2({\bf x})
=S_3({\bf x})=0$. Since $S_0 =S_1^2+S_2^2+S_3^2$, this condition
gives a zero intensity point where polarization is not well defined.

Performing such an analysis on the incident, reflected and refracted
beams, we found extended regions where the reflected and refracted
beams are approximately linearly polarized if the incident beam is
circularly polarized in the sense that ${\cal E}\simeq \pm i{\cal
B}$; this is illustrated in Figs.~(4-6). As expected, the
polarization of the ordinary beam is orthogonal to that of the
extraordinary beam. The behavior of the polarization singularities
is quite interesting: for the incident beam, an optical vortex is
located at its center so that ${\bf E({\bf 0})\times \hat e_q} ={\bf
0}$. But neither the refracted nor the reflected beams are null at
that point: this means that a polarization vortex may not survive
the reflection and refraction process. However, $S_0=0$  at other
locations both at the interface and along the produced beams and, in
fact, the number of these singularities increases as the refracted
beams evolve. This phenomenon is similar to that described in
Ref.~\cite{flossmann,flossmann2} for a Laguerre-Gaussian beam.
\begin{figure}
\includegraphics[scale=0.3]{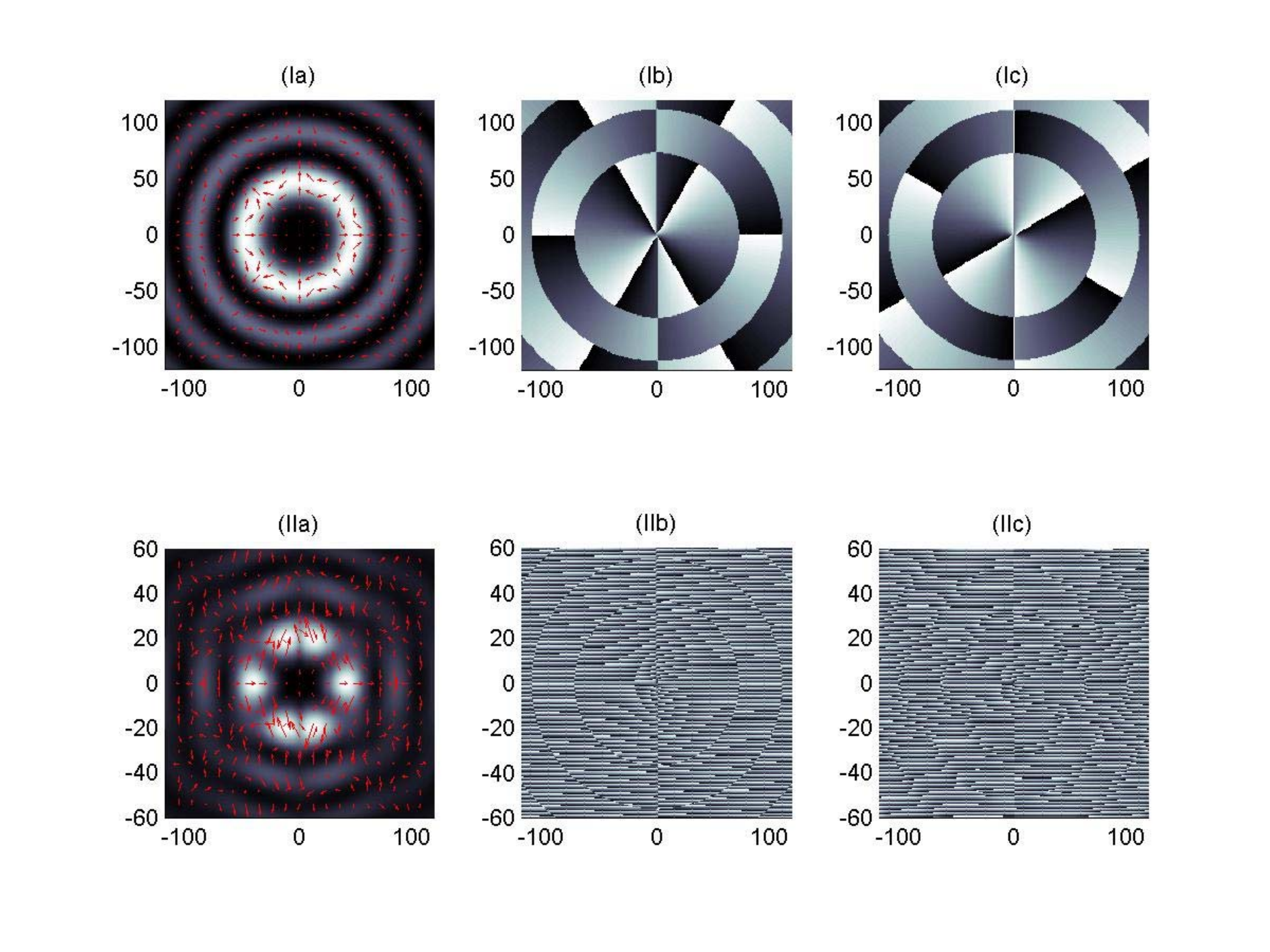}
\caption{(Color on line) Intensity, polarization  and phase
patterns for a Bessel beam of order $m=2$ with ${\cal B}= i{\cal E}$
and $\zeta = \pi/36$. The patterns are produced in a plane
perpendicular to the main axis of propagation (upper row) and at an
angle $\alpha= \pi/3$ with that axis (lower row). The second and
third columns show the density plot of the phase for the electric
field along two perpendicular vectors on the corresponding plane.
The brighter (darker) regions correspond to phases close to $\pi$
($-\pi$). The unit of length is $\lambda = c/\omega$. } \label{fig4}
\end{figure}

\begin{figure}
\includegraphics[scale=0.3]{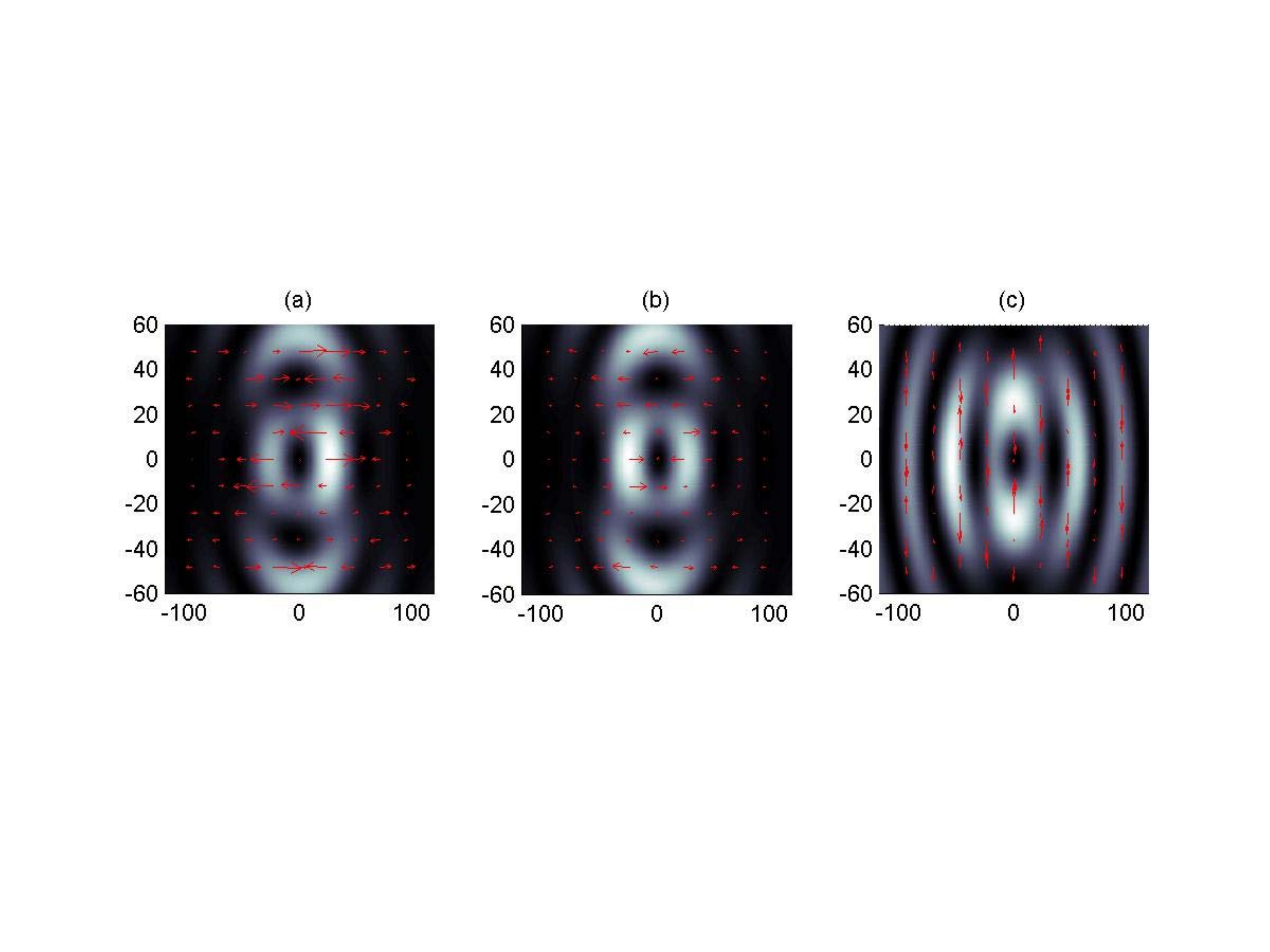}
\caption{(Color on line) Intensity and polarization patterns for the
reflected, ordinary and extraordinary beams at the interface
resulting from an incident Bessel beam of order $m=2$ with the same
configuration as described in Fig.~1. Notice the different scales.
The observation window dimensions are given in units of $\lambda =
c/\omega$. } \label{fig5}
\end{figure}

Given the analytic expressions, the phase patterns can be easily
analyzed for each component of the electric field. Such an analysis
is essential for the understanding of the mechanical properties of
the beams. For a Bessel beam of order $m$, Eq.~(\ref{1_a}) implies
that the components $E_i$ are superpositions of fields with phase
$m^\prime\phi + k_zz$, where $m^\prime = m,m\pm 1$. The linear
momentum density along the $z$ direction is proportional to $k_z$
and the term proportional to $\phi$ determines the orbital angular
momentum density along  that axis, which is given by the formula
\begin{equation}
{\cal L}_z = \frac{1}{4\pi}\sum_{i=x,y,z}
 E_i[\frac{\partial}{\partial \phi}]A_i.
\end{equation}
Accordingly, phase vortices with a topological charge $m$ contribute
to the orbital angular momentum with a factor proportional both to
$m$ and to the moduli of the corresponding electric field components.
For reference, the phase patterns of a $m=2$ Bessel beam
 for two perpendicular components of
the electric field in two distinct planes are shown in
Fig.~(\ref{fig4}): one perpendicular to the main axis of propagation
(first row) and another at an angle $\alpha=\pi/3$ with the former
(second row). In the first row, the standard $m^\prime = 1,2,3$
optical charge vortices are observed, but in the second row the
phase patterns exhibit an extremely complex topological structure
(dislocations, vortices and bifurcations) on both short and long
length scales of order $1/k_z$ and $1/k_\bot$ respectively. It is
also worth noticing that the electric field perpendicular to the
tilted plane is not negligible.

\begin{figure}
\includegraphics[scale=0.3]{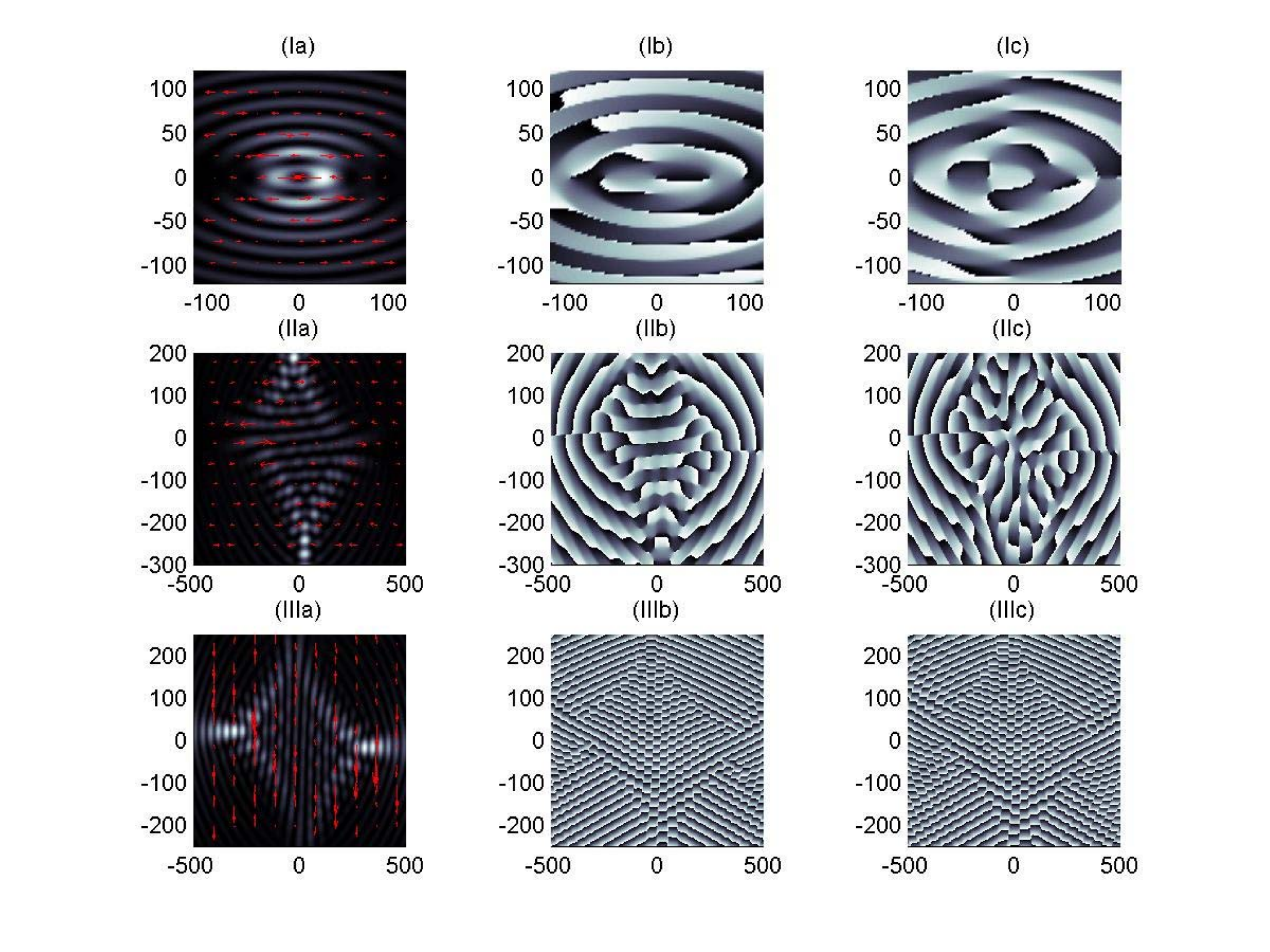}
\caption{(Color on line) Intensity, polarization and phase patterns
for the reflected (1st row), ordinary (2nd row) and  extraordinary
(3rd row) beams resulting from an incident Bessel beam of order
$m=2$ with the same configuration as described in Fig.~1. The plane
of observation is perpendicular to the main axis of propagation of
each beam and is located at a distance $10500\lambda$ from the
interface plane. The brighter (darker) regions correspond to phases
close to $\pi$ ($-\pi$). Unit of length is $\lambda = c/\omega$.
Notice the different coordinate scales.} \label{fig6}
\end{figure}

The reflected, ordinary and extraordinary beams just at
the interface posses, as expected, a phase structure
qualitatively similar to that shown in Fig.~4 in the (IIb) and (IIc)
plots. In particular, the reflected wave exhibits an elliptic-like
symmetry with phase vortices in the plane perpendicular to its axis;
this is shown in the first row of Fig.~(\ref{fig6}).

Due to the anisotropy inside the crystal, the only component of the
angular momentum that can be conserved is the one along the crystal
axis \cite{a-m-i}; its density is given by
\begin{equation}
{\cal L}_z = \frac{1}{4\pi}\sum_{i=x,y,z}
 D_i[\frac{\partial}{\partial \phi}]A_i.
\end{equation}
where ${\bf D}$ is the electric displacement vector. If the ordinary
and extraordinary beams do not propagate along the crystal axis,
their angular momentum along their propagation axis is not
conserved. This in turn should be manifested as the creation and
annihilation of phase vortices as the beams propagate in the
crystal. In the second and third rows of Fig.(\ref{fig6}) the phase
diagrams for the ordinary and extraordinary beams are illustrated at
planes located deep inside the crystal: both beams have a rich
topological structure but the characteristic length is much smaller
for the extraordinary wave. Notice that for both ordinary and
extraordinary waves, there is a strong correlation between the phase
diagrams and the intensity patterns.

\section{Conclusions}

In this work, explicit formulas have been obtained that permit an
analytic treatment of birefringence in uniaxial crystals. These
expressions are valid for arbitrary orientations of the incident
beam and the axis of the crystal with respect to the plane defining
the interface.  An illustrative application has been worked out in
detail for a Bessel beam and a particular orientation of the crystal
axis: it has shown that the electric fields of ordinary,
extraordinary and reflected waves take a relatively simple analytic
form in terms of a circuit integral. With these results, it was
shown that the polarization and phase vortices of the ordinary and
extraordinary beams evolve into a complex structure for sufficiently
thick crystals. This structure is quite different for the ordinary
and extraordinary beams and necessarily leads to different
mechanical properties of the field. In was also shown that the
reflected beam has elliptic symmetry, exhibits phase and polarization
vortices and is invariant under propagation. We expect these results
to be useful for the correct characterization of general waves in
birefringent crystal as they are widely used at present in optical
experiments.

\end{document}